\documentclass[referee]{raa}            
\usepackage{txfonts}
\usepackage{graphicx,times}             

\newcommand{\kms}{km\,s$^{-1}$}

\begin{document}

   \title{LAMOST Medium-Resolution Spectroscopic Survey (LAMOST-MRS): Scientific goals and survey plan
}

   \volnopage{Vol.0 (200x) No.0, 000--000}      
   \setcounter{page}{1}          

   \author{Chao Liu
      \inst{1,2}
   \and Jianning Fu
      \inst{3}
   \and Jianrong Shi
      \inst{4,2}
   \and Hong Wu
      \inst{4}
   \and Zhanwen Han
      \inst{5}
   \and Li Chen
      \inst{6,2}
   \and Subo Dong
      \inst{7,8}
   \and Yongheng Zhao
      \inst{4,2}
   \and Jian-Jun Chen
      \inst{4}
   \and Haotong Zhang
      \inst{4}
   \and Zhong-Rui Bai
      \inst{4}
   \and Xuefei Chen
      \inst{5}
   \and Wenyuan Cui
      \inst{9}
   \and Bing Du
      \inst{4}
   \and Chih-Hao Hsia
      \inst{10}
   \and Deng-Kai Jiang
      \inst{5}
   \and Jinliang Hou
      \inst{6,2}
   \and Wen Hou
      \inst{4}
   \and Haining Li
      \inst{4}
   \and Jiao Li
      \inst{5,1}
   \and Lifang Li
      \inst{5}
   \and Jiaming Liu
      \inst{4}
   \and Jifeng Liu
     \inst{4,2}
   \and A-Li Luo
      \inst{4,2}
   \and Juan-Juan Ren
      \inst{1}
   \and Hai-Jun Tian
      \inst{11}
   \and Hao Tian
      \inst{1}
   \and Jia-Xin Wang
      \inst{3}
   \and Chao-Jian Wu
      \inst{4}
   \and Ji-Wei Xie
      \inst{12,13}
   \and Hong-Liang Yan
      \inst{4,2}
   \and Fan Yang
      \inst{4}
   \and Jincheng Yu
      \inst{6}
   \and Bo Zhang
      \inst{3,4}
   \and Huawei Zhang
      \inst{7,8}
   \and Li-Yun Zhang
      \inst{14}
   \and Wei Zhang
      \inst{4}
   \and Gang Zhao
      \inst{4}
   \and Jing Zhong
      \inst{6}
   \and Weikai Zong
      \inst{3}
   \and Fang Zuo
      \inst{4,2}
   }

   \institute{Key Lab of Space Astronomy and Technology, National Astronomical Observatories, Chinese Academy of Sciences, Beijing 100101, China; {\it liuchao@nao.cas.cn}\\
   	\and
   	University of Chinese Academy of Sciences, 100049, China\\
   	\and 
   	Department of Astronomy, Beijing Normal University, Beijing, 100875, China\\
   	\and 
   	Key Lab of Optical Astronomy, National Astronomical Observatories, Chinese Academy of Sciences, Beijing 100101, China \\    
   	\and
   	Yunan Astronomical Observatory, China Academy of Sciences, Kunming, 650216, China\\
   	\and
   	Key Laboratory for Research in Galaxies and Cosmology,   Shanghai Astronomical Observatory, Chinese Academy of  Sciences, 80 Nandan Road, Shanghai 200030, China\\
   	\and
   	Department of Astronomy, School of Physics, Peking University, Beijing 100871, China\\
   	\and
   	Kavli Institute of Astronomy and Astrophysics, Peking University, Beijing 100871, China\\
   	\and
   	Department of Physics, Hebei Normal University, Shijiazhuang, 050024, China\\
   	\and
   	Space Scinece Institute, Macau University of Science and Technology, Taipa, Macau\\
   	\and
   	Center for Astronomy and Space Sciences, China Three Gorges University, Yichang, 443002, China\\
   	\and
   	School of Astronomy and Space Science, Nanjing University, Nanjing 210023, China\\
   	\and
   	Key Laboratory of Modern Astronomy and Astrophysics, Ministry of Education, Nanjing 210023, China\\
   	\and
   	College of Physics/Department of Physics and Astronomy, Guizhou University,
   	Guiyang 550025, P.R. China\\
}

   \date{Received~~; accepted~~}

\abstract{ Since September 2018, LAMOST starts a new 5-year medium-resolution spectroscopic survey (MRS) using bright/gray nights. We present the scientific goals of LAMOST-MRS and propose a near optimistic strategy of the survey. A complete footprint is also provided. Not only the regular medium-resolution survey, but also a time-domain spectroscopic survey is being conducted since 2018 and will be end in 2023. According to the detailed survey plan, we expect that LAMOST-MRS can observe about 2 million stellar spectra with $R\sim7500$ and limiting magnitude of around $G=15$\,mag. Moreover, it will also provide about 200 thousand stars with averagely 60-epoch observations and limiting magnitude of $G\sim14$\,mag. These high quality spectra will give around 20 elemental abundances, rotational velocities, emission line profiles as well as precise radial velocity with uncertainty less than 1\,km\,s$^{-1}$. With these data, we expect that LAMOST can effectively leverage sciences on stellar physics, e.g. exotic binary stars, detailed observation of many types of variable stars etc., planet host stars, emission nebulae, open clusters, young pre-main-sequence stars etc.
}

   \authorrunning{Liu et al.}            
   \titlerunning{LAMOST-MRS scientific goals and survey plans}  

   \maketitle

%
%
\section{Introduction}           
\label{sect:intro}
LAMOST (Large Sky Area Multi-Object fiber Spectroscopic Telescope; also known as Guo Shou Jing telescope) is a 4-meter quasi-meridian reflective Schmidt telescope with 4000 fibers installed on its 5-degree-FoV focal plane. These  configurations allow it to observe spectra for at most 4000 celestial objects simultaneously (\cite{2012RAA....12.1197C, 2012RAA....12..723Z}). From 2011 to 2018, it has achieved its first stage low spectral resolution ($R\sim1800$) survey and obtained more than 10 million spectra, including about 9 million stellar spectra \footnote{http://dr5.lamost.org}. As one of the largest spectra database,  LAMOST survey has shown substantial impact in diverse fields, from stellar astrophysics (\cite{2014ApJ...788L..37G, 2015ApJ...807....4L, 2016NatCo...711058K, 2016ApJ...818..202L, 2017RAA....17...10W, Ren2018, 2019MNRAS.490..550L, 2019Natur.575..618L, 2019ApJ...872L..20G}),
the Milky Way science (\cite{2015ApJ...809..145T, 2015RAA....15.1209X, 2016MNRAS.463.2623H, 2017RAA....17...96L, 2017ApJ...835L..18L, 2017RAA....17..114T, 2017MNRAS.470.2949W, 2018ChA&A..42....1L, 2018ApJS..238...16L, 2018ApJ...865L..19T, 2018MNRAS.478.3367W, 2018MNRAS.473.1244X, 2018MNRAS.475.1093Y, 2018ApJ...868..105Z, 2019ApJ...874..138L, 2019ApJ...871..184T, 2019ApJ...877L...7W, 2019MNRAS.482.2189W}),
to external galaxies and QSOs (\cite{2016RAA....16...43S, 2010RAA....10..737W, 2010RAA....10..745W, 2019ApJL...884...7H}).

The 16 spectrographs installed on LAMOST has been upgraded during 2017 and currently support not only spectral resolutions of $R\sim1800$, which is the same as the first stage survey, but also the new medium resolution with $R\sim7500$. All spectrographs can be switched between the two modes during a day time to adopt different surveys. 

At $R\sim7500$, the blue and red cameras located on each spectrograph cannot cover wavelength ranges as large as the low resolution. Instead, the blue cameras can only cover the wavelength range from 4950\,$\AA$ to 5350\,$\AA$, while the red cameras cover from 6300\,$\AA$ to 6800\,$\AA$. With medium-resolution spectra, one can achieve the precision of the radial velocities to around 1\,\kms\ for late type stars. The rotation of the stars can be well detected with uncertainty of around 10\,\kms\ with the medium-resolution spectra. And the coverage of the wavelength enables to measure about 20 elemental abundances, including Li, C, Na, Mg, Si, Ca, Sc, Ti, V, Cr, Mn, Fe, Co, Ni, Cu, Ba, Y, Sm, and Nd. In the mean time, the red part of the spectra can well cover H$\alpha$, [NII], and [SII] emission lines, which is crucial for the science related to the Galactic nebulae, such as the HII regions, the supernova remnants, the planetary nebular etc. The H$\alpha$ profile is also very important in research on the proto-planetary disk surrounding the very young stars. 

Since October 2018, LAMOST started the second stage survey program, LAMOST II, which contains both low- and medium-resolution spectroscopic surveys. LAMOST II will take around 50\% nights (dark/gray nights) to continue the previous low-resolution spectroscopic survey. The other 50\% nights (bright/gray nights) are assigned for medium-resolution survey. While the scientific goals of the LAMOST II low-resolution survey is generally the same as the first 5-year survey (Deng et al. 2012), the new medium-resolution survey (hereafter MRS) will aim for new goals, including Galactic archaeology, stellar physics, star clusters, star formation, exoplanet host stars, Galactic nebulae etc.

Among these scientific goals, synergy with other international space-based surveys can manifest the impact of the MRS survey. 
In the past few years, {\it Kepler} (\cite{2010Sci...327..977B}) and K2 missions have provided accurate time-domain photometric data for hundreds of thousands of stars distributed in lots of fields in the sky. These data are invaluable legacy in many topics from exoplanet to asteroseismology. As its successor, NASA Transiting Exoplanet Survey Satellite (TESS, \cite{Ricker14})
launched in April 2018 and will conduct a two-year survey of nearly the entire sky to search for planets around bright main-sequence stars, potentially yielding thousands of new exoplanet discoveries including those smaller than the Neptune. In the mean time, the time-domain data are also extremely critical for asteroseismology, pulsators, and binary stars. Ground-based spectroscopic data for a large number of {\it Kepler}, {\it K}2, and TESS observed stars will remarkably extend the science of these missions and are expected to substantially amplify the science impacts.

The \emph{Gaia} satellite has been launched by ESA since December 2013 and the \emph{Gaia} DR2 data have been publicly released since April 25 2018 (\cite{2016A&A...595A...1G,  2018A&A...616A...1G}). The final data release will be expected to come out in 2022. As an ambitious mission, \emph{Gaia} has provided extremely accurate astrometry as well as photometry for 1.7 billion stars with $G$ band magnitude brighter than $20.7$\,mag\footnote{https://www.cosmos.esa.int/web/gaia/release}. In the next decades, \emph{Gaia} will be the one of the most important data legacy in almost all fields of astronomy.


The high-cadence high-precise time-domain photometric data, extremely precise trigonometric parallaxes and proper motions combined from these space missions will be perfect supplementary of the LAMOST-MRS survey, which can provide precise radial velocities, stellar rotations, atmospheric parameters, elemental abundances, and time-domain spectroscopic data. 

In this paper, we propose the scientific goals of the LAMOST-MRS and the targeting and survey plans. Section 2 describes the proposed scientific goals. Section 3 provides a brief report of the test observing with mid-resolution spectrographs and a simulation of the survey to find the approximately optimistic solution of the survey such that all scientific goals can be satisfied. Section 4 gives the detailed sky footprint and targeting plan. And Section 5 gives a brief summary.  

\section{Scientific goals}
\label{sect:sci}
As a stellar medium-resolution spectroscopic survey for millions of stars, LAMOST-MRS aims at diverse scientific goals, including
\begin{itemize}
\item Binarity/Multiplicity,
\item Stellar pulsation,
\item Star formation,
\item Emission nebulae,
\item Galactic archaeology,
\item Host stars of exoplanets,
\item Open clusters.
\end{itemize}
In the next sub-sections, we briefly introduce these scientific goals one by one.

\subsection{Binarity}\label{sec:binary}
More than 50 percent of stars are in binary or multiple systems in the Milky Way (Duchene \& Kraus 2013). The binary interactions can dramatically affect the evolutionary tracks of both components in binary systems. Therefore the binary evolution, even in main-sequence stage, can be significantly different from single-star evolution (\cite{Jones2017}). 

A number of outstanding scientific questions are related to binary stars. Massive binaries containing O or B-type stars are linked to the black hole/neutron star binaries, which are considered as sources of gravitational wave (\cite{Abbott2016, DeMarco17}). Binaries composed of a white dwarf and a A, F, G or K type main-sequence star or composed of two white dwarfs are likely the progenitors of cosmologically important type Ia supernovae (\cite{Webbink1984, 2004MNRAS.350.1301H}). Meanwhile, it has been widely known that binary evolution produces most exotic stellar objects e.g. hot subdwarf stars, barium stars, symbiotic stars, double degenerates, cataclysmic variables, AM CVn stars, binary millisecond pulsars, bipolar planetary nebulae, extremely low-mass white dwarfs, etc (\cite{1995MNRAS.272..800H, 1995MNRAS.277.1443H, 1998MNRAS.296.1019H, 2002MNRAS.336..449H, 2004MNRAS.350.1301H, Chen2009, 2013MNRAS.434..186C, 2017MNRAS.467.1874C, 2019ApJ...871..148L}). However, several crucial processes in binary evolution such as mass and angular momentum loss, dynamical mass transfer and common envelope evolution etc. have not been well understood yet. These processes determine the formation of the exotic objects and the ultimate fate of a binary. Searching for some of the exotic objects (they are actually binaries in various evolutionary stages) and determining their physical properties are greatly helpful in understanding the important processes in binary evolution, then the formation of type Ia supernovae and sources of gravitational waves.

In addition to the highly interesting exotic binaries, the statistical properties of the binaries in the field as well as in stellar systems (open clusters, globular clusters, associations etc.) are also of importance in the sense that they can provide critical information of the formation and evolution of the binary systems. The fraction of binaries in a stellar population can vary with metallicity (\cite{2010ApJS..190....1R, 2014ApJ...788L..37G, 2019MNRAS.482L.139E, 2019ApJ...875...61M, 2019MNRAS.490..550L})
and the mass of primary stars (\cite{2010ApJS..190....1R, 2013ARA&A..51..269D}).
In the mean time, the distribution of the mass-ratio of binaries are also different in binary systems with different primary mass (\cite{2013ARA&A..51..269D,2019MNRAS.490..550L}).
The distribution of the period of binaries in the solar neighborhood is unveiled by various works (\cite{1991A&A...248..485D, 2010ApJS..190....1R}).
It is not clear, however, whether this period distribution is universal. It is also found that the orbital eccentricity is averagely small at small period, implying that most of the orbits of close binaries have been circularized. According to various simulations (\cite{2008ApJ...681..375K, 2009MNRAS.399.1255M, 2010ApJ...708.1585K, 2014MNRAS.439.1884T, 2018ApJ...854...44M}),
it seems that, the close binaries are formed from disk fragmentation, which can lead to the anti-correlation between the binary fraction and metallicity, while some observational evidence hints that the wide binaries may have different channel of formation (\cite{2019MNRAS.482L.139E,HJT2019,2019MNRAS.489.5822E}).
Moreover, it has been 
showed that, at the early stage when the field stars were still staying in the embedded clusters in which they were born, the dynamical process has already effectively changed the statistical properties of the binaries by stripping out the smaller and distant companions of the binaries (\cite{2011MNRAS.417.1702M,2019MNRAS.490..550L}). 

With LAMOST-MRS survey, we expect to discover more exotic binaries under different evolutionary stages, such as CVs, Hot subdwarf stars, WD+MS binaries, symbiotic stars, barium stars, the PNe with binary central stars, etc. Medium-resolution spectra with accurate measurement of the radial velocities enables to derive their physical and orbital parameters in high precision. These are crucial in understanding common-envelope phase, the formation of SN Ia, and gravitational wave sources. We can more precisely obtain the orbital parameters of eclipsing binaries combined by the time-domain photometric light curves and LAMOST radial velocity curves. 

We can also determine the behavior (single or double emissions, absorption) of the H${\alpha}$ lines. Using LAMOST MRS time-domain data, we can study the time or orbital phase evolution of the H${\alpha}$ lines, especially short-time scale variation. It is possible for late-type eclipsing binaries to detect the plage or flare events from H${\alpha}$ line variation. We can compare the activity fraction of single or binary system and discuss the effect by the close companion (\cite{2012AJ....144...93M}). Starspots and energy of flares derived from time-domain photometric data will be complementary to the spectroscopic features (\cite{zhang2020}). Finally, we can determine the relationship of photometric and chromospheric activity for late-type eclipsing binaries. This will promote the exact mechanism behind the magnetic field generation of late-type eclipsing binaries.

Meanwhile, the multi-epoch MRS survey for a large number of stars with different spectral types allows us to accurately count the binary fraction, mass-ratio distribution, and orbital eccentricity distribution for close binaries with period smaller than around one thousand days. In fact, for many of these stars, we are able to partly solve their orbital parameters from their radial velocity curves. Moreover, we expect that a fraction of these spectroscopic binaries are double-line binaries (SB2), which orbital parameters can be completely solved out from their radial velocity curves. As a by-product, we are able to provide the accurate stellar mass for a few thousands of stars with different spectral types.

Combined with time-domain photometric data, such as TESS, {\it Kepler}, and ZTF (\cite{2019PASP..131a8003M}), more information can be provided especially for the eclipsing binaries. These will be a large amount of invaluable data legacy for the studies of binaries.

\subsection{Variable stars}\label{sec:varible}

Stellar pulsations provide the unique technique for one to probe the internal structures and chemical compositions of stars by means of asteroseismology (\cite{aerts10}). The well-defined internal physics of stars may inversely offer precise information to constrain stellar evolutionary theory (e.g., \cite{giammi18}). Currently, most of the asteroseismic data are purely from high-precision high-cadence time-domain photometric data, such as {\it Kepler} and TESS. However, lack of the precise measurement of the stellar atmospheric parameters, including the effective temperature, surface gravity, metallicity, and rotation, will limit many studies in asteroseismology. Successful experience from past years suggests that LAMOST is an ideal instrument for follow-up observations of those spaceborne surveys (e.g.,\cite{DeCat15,Zong2018})


With the LAMOST-MRS, precise atmospheric parameters will be obtained for various types of stars, which will reduce the parameter space when searching for the optimal seismic models, in particular for large amplitude pulsators. The time-series variations of radial velocity from LAMOST-MRS, together with the light curves derived from the high-precision photometry, will offer better constraints on the determination of stellar parameters, including the dynamical processes of pulsating stars as well as the masses and radii
of the components for binaries (\cite{Murphy2018}).

LAMOST-MRS will help to provide an independent measurement of the rotation periods of stars which have been also estimated from the {\it Kepler} photometry. It is also helpful to determine the intensity of stellar activity on the surface using the particular emission lines.

These topics can get benefit from time-domain spectroscopic observations for a large amount of target stars, especially when high-precision long-time-coverage light curves, e.g. the observations from {\it Kepler} and {\it K}2 campaigns, are available. 

\subsection{Star forming regions}\label{sec:sfr}
The detailed process of star formation is far from clear. In general, the young stars are mostly born in clusters embedded in molecular clouds (\cite{2003ARA&A..41...57L}).
 After a fairly short time scale, most of the young embedded clusters are disrupted and stars are scattered into the field. It is of high interests that how young stars co-evolve with the molecular clouds and how they are decoupled (\cite{2016A&A...590A...2S}). Another interesting question is the binary fraction in the young stellar populations. The comparison between the very young and old stellar populations is important to understand whether the binary stars have been experienced substantial dynamical processing during their stay in the embedded clusters (\cite{2011MNRAS.417.1702M, 2019MNRAS.490..550L}).
Young low-mass stars may still keep a weak proto-planetary disk surrounding the star. This can be observed from the emission of the H$\alpha$ line. The variation of the emission line profiles can provide detailed properties of these disks. Moreover, the evolution of the Lithium in the atmosphere of young stars is also an interesting open question, which is related to the cosmological evolution of Lithium.

LAMOST-MRS will cover a few nearby star forming regions, such as Taurus, Orion, Perseus, Gemini etc., and a dozen of star associations. We will sample these regions and obtain medium-resolution spectra for both massive and low-mass young stars. It is possible to observe the late stage of the proto-planetary disks in optical band with LAMOST-MRS spectra. Moreover, because the LAMOST-MRS spectra contains Li line at 6707$\AA$, it allows us to approximate the evolution of the Lithium in the young stars. 

\subsection{Galactic Archaeology}\label{sec:chem}

It is of great importance to unravel the formation and evolutionary history of the Milky Way {key of which includes reconstructing the lost stellar substructures of the early Galaxy. Although impossible to identify them from the spatial distributions, as stellar abundances reflect the chemical state of the gas where stars have been born, remnants of the disrupted yet now-dispersed ancient building blocks, e.g. dwarf galaxies and globular clusters, shall be identified through exploring their stellar chemistry} although it is impossible to identify them from the spatial distributions. 
Kinematic properties can be a critical supplementary to chemical tagging when one take into account the accurate astrometry from \emph{Gaia} catalog. Chemical-kinematic tagging will finally provide an comprehensive view of the ensemble of the Milky Way in the context of the near field cosmology.

The LAMOST-MRS will observe around a million medium-resolution spectra for FGK stars. These enables us to measure around 20 elemental abundances, including light proton-capture elements, $\alpha$-elements, odd-Z elements, iron-peak elements, and neutron-capture elements, in reasonable precision (\cite{2019arXiv191013154Z,2020ApJS..246....9Z}). {which would be sufficient to represent the main nucleosynthetic processes in both dwarf and giant stars, and allow chemical tagging of individual stars to their original formation event.}
Combining with the radial velocities with uncertainty of around 1\,km s$^{-1}$ and \emph{Gaia} astrometry, which will permit differentiation of stars by their kinematics,
we {will be able to reconstruct the original stellar building blocks, and to investigate the history of the formation and evolution of the Galaxy
by means of chemical-kinematic tagging.

\subsection{Emission nebulae}\label{sec:neb}

The emission nebulae include planetary nebulae (PNe), HII regions, supernovae remnants (SNRs), Herbig-Haro (HH) objects etc (\cite{wu2020}). These objects have various scientific interests covering the nature of the ionized gas, star formation and the evolution of the Milky Way.

The three-dimensional ionized structures of PNe reflect various nebular expansion velocities. The interesting scientific questions on PNe include 3D ionized structures and electron density distributions of Galactic PNe, the evolution of various morphological PNs, and the chemical abundance of our Galaxy traced by PNe.

\hii\ regions are gas nebulae ionized by high-energy radiation (mostly are ultraviolet photons) from young massive stars. 
As the distribution of these stars and the surrounding gas are irregular, in most situation, the shape of the \hii\ regions is far from simple Stro\"mgren sphere, instead, they often appear clumpy and filamentary (\cite{2011MNRAS.413.2242M}). 
\hii\ regions are natural laboratories for studying star-forming actives and interactions between the massive stars and surrounding ISM (\cite{2018ApJ...853..151M}). Meanwhile, the spatial variation of chemical abundances across the Galactic disk can also be well measured with HII regions (\cite{1995ApJ...444..721S,1997ApJ...478..190A,2000MNRAS.311..329D,2005ApJ...618L..95E,2006ApJS..162..346R,2011ApJ...738...27B,2017MNRAS.471..987E,2017A&A...597A..84F}).

SNRs are the material ejected in an explosion of SN, which can continue their life through interaction with the surrounding interstellar medium (ISM) for thousands or even up to a million year. They are important sources of the injection of energy and heavy elements into the interstellar medium, and are believed to be the possible sources of the Galactic cosmic rays. Moreover, SNRs probe the immediate surroundings of supernovae, shaped by their progenitors. 
The main features of SNRs are strong shock waves, amplified magnetic field, ultra-relativistic particles (cosmic rays) generation and associated synchrotrion radiation (\cite{Arbutina2017}). 
HH objects (\cite{Ambartsumian1954}) are formed when narrow jets of partially ionized gas ejected by stars collide with nearby gas and dust at speeds of several hundred kilometers per second. 
They are small-scale shock regions intimately associated with star forming regions (\cite{Schwartz1983}) and can trace supersonic shock waves (\cite{bally2016}), often found in large groups, and some can be seen around a single star, aligned with its rotational axis. 

LAMOST-MRS will be used as a super large IFU when observe a large area of the emission nebulae, and to provide high quality medium-resolution spectra containing most of the prominent emission lines e.g. H$\alpha$, [NII], [SII], [OIII] etc. The emission nebulae can be well discriminated from the log(H$\alpha$/[\ion{N}{ii}]) vs. log(H$\alpha$/[\ion{S}{ii}]) (S2N2) diagnostic diagram (\cite{2014A&A...563A..63H}). LAMOST-MRS will obtain large number of optical spectra of \hii\ regions in the Galactic disk, which allow us to study the radial abundance gradient and to re-check whether there is evidence of flaring at the outer disk (\cite{Wang2018}). We will focus on a dozen of largely extended SNRs and efficiently map their details with the help of the wide field-of-view, multi-fibers, and high-precision in emission line profiles provided by LAMOST-MRS (\cite{Ren2018}). LAMOST-MRS will provide opportunity to study the most of known HH objects in details and identify more HH objects candidates.

\subsection{Host Stars of Exoplanet and Synergy with TESS}\label{sec:exoplanet}

Both {\it Kepler} and TESS are imaging surveys aiming to exoplanets detection from transit method. For the transit method, the accurate knowledge of the host stars is crucial to characterize the exoplanets and study the dependence of their distribution on the stellar environment. The data of the LAMOST-MRS survey will offer a great synergy with TESS by providing a large and complete spectroscopic sample of exoplanet host stars as well as the parent control samples for population studies.

Previous studies (see, e.g., \cite{Dong14,Xie16,Wang16}) have demonstrated that the stellar parameters (\cite{Luo15}) derived from LAMOST low resolution spectroscopic observations of the {\it Kepler} field (\cite{DeCat15,Ren16}) reach high precision for main sequence stars ($\sigma_{[Fe/H]} \approx 0.1$\,dex,  $\sigma_{\log(g)} \approx 0.1$\,dex). The large and unbiased LAMOST-{\it Kepler} sample with well characterized stellar properties have led to many new insights into {\it Kepler} planets and their distributions, including (not limited to) the eccentricity distribution of {\it Kepler} planets (\cite{Xie16}), discovery of a new population of exoplanets with key properties similar to the hot Jupiters (\cite{Dong18}) and a constraint on the fraction of stars with {\it Kepler}-like systems (\cite{Zhu18}). With  LAMOST-MRS and the new synergy with TESS, we expect that LAMOST data will continue to contribute significantly in the following two main additional aspects of exoplanet study.

As stars are commonly formed in binary systems, studying planets in binaries is important to understand the general process of planet formation. However, existing observations of planets in binaries are incomplete and biased. Radial velocity (RV) exoplanet surveys usually avoid binaries in the initial target selection. Although at the target selection stage, transit surveys like {\it Kepler} and TESS are not particularly biased against binaries, transit observations alone typically only yield limited constraints on binaries from detecting eclipses for binary companions at relatively close separations with orbital planes well aligned with the line of sight. High-resolution direct-imaging follows up of transit targets with adaptive optics have revealed that a significant portion exoplanets hosts are wide binaries (e.g., \cite{Law14}). LAMOST-MRS can achieve a radial velocity precision of $\sim 1$\,km/s, and in complementary with high-resolution imaging observations, LAMOST-MRS RV follow-up data can reveal binary companions at close and intermediate separations at nearly all orbital orientations. 

	TESS is an all-sky survey, which can potentially find exoplanets across various stellar types and populations. With the large samples accessible to LAMOST-MRS and additional astrometric information from {\it Gaia}, it is also possible to probe the Galactic distribution of planets (thin disk vs. thick disk) using metallicity and kinematics information. In particular, LAMOST-MRS could measure the abundances of more element than LAMOST low-resolution data, e.g., $\alpha$ elements [$\alpha$/Fe], which can aid the stellar population diagnostic (\cite{2019arXiv191013154Z}). Based on our current knowledge of the structure of thin and thick disks (e.g., \cite{Bland-Hawthorn16}), we expect that $\sim 10$\% of our targets are thick-disk stars, and the LAMOST-MRS TESS synergy may find $\sim 100$ or more planet systems around thick-disk stars. Statistical studies by comparing the exoplanet systems of different star populations will provide new insight into planet formation during the history of the Galaxy evolution. 


\subsection{Open clusters}\label{sec:oc}

Open clusters (OCs) have long been used to trace the structure and evolution of the Galactic disk. Open clusters have large age and distance spans and can be relatively accurately dated; the spatial distribution and kinematical properties of OCs provide critical constraints on the overall structure and dynamical evolution of the Galactic disk. Meanwhile, their [M/H] values serve as excellent tracers of the abundance gradient along the Galactic disk, as well as many other important disk properties, such as the age-metallicity relation (AMR), abundance gradient evolution, etc. 

With the great ability of LAMOST-MRS, we expect to survey dozens of open clusters to obtain stellar radial velocities as well as abundance information of around 3  $\times$ 10$^{5}$ stars complete to $G\sim15$\,mag to cover the cluster central areas as well as their surrounding fields. This will give the largest magnitude-complete spectroscopic dataset for studying the Milky Way open clusters. Meanwhile, the observations for total of 3 $\times$ 10$^{5}$ stars nearly uniformly distributed around the open clusters will provide most extending and well-representative cluster-halo sample with spectroscopic data, which would be of great importance for probing the overall spatial structure and dynamical evolution of open clusters (e.g., \cite{2019A&A...624A..34Z}).

The large amount of up-to-date homogeneous open cluster data from LAMOST-MRS will lead to the most reliable membership determination for sample clusters using accurate radial velocity, combining with the most precise proper motion data from Gaia DR2. These will significantly purify the color-magnitude diagrams of sample open clusters, obtaining the essential parameters of clusters, such as distances and ages, and also provide the best basis for investigating internal kinematic properties of open clusters.

Our target samples contain a large portion of young (log t $\le $ 7.5) clusters. Young open clusters, which retain a representative mass spectrum, are favorable targets to determine the IMF. However, in the absence of spectroscopic data IMF determinations are subject to uncertainties in the slope measurements and contamination. Meanwhile, only limited spectroscopic surveys to young clusters have been conducted to date. 

From LAMOST-MRS survey, based on most reliably kinematic-based membership estimation and together with precise Gaia DR2 photometric data, comprehensive photometry and spectroscopy of hundreds of members in each of young clusters will be obtained. This will result in unprecedented improvements in the study of IMF as well as the formation and early revolution of star clusters.


\section{The survey strategy and plan}
\label{sec:obscon}
In this section, we present the expected survey performance, survey strategy and plan.

\subsection{Expected performance}\label{sec:perf}
We have been conducting a commissioning survey between October 2017 and June 2018 and obtained about 5 million single exposure medium-resolution stellar spectra. We tested different exposure time from 600 to 1800\,s and finally decide to take 1200\,s as the default exposure time, which can provide a balance between the signal-to-noise ratio (S/N) of single exposure and limiting magnitude. Figure~\ref{fig:Gsnr} shows how S/N of the medium-resolution spectra changes with \emph{Gaia} $G$-band magnitude for around 5 million 1200\,s exposure spectra  observed during the commissioning survey. 
The left panel shows that, in the blue band, the S/N is larger than 5 when $G<14.5$\,mag. In the right panel, the limiting magnitude can go down to 15\,mag when S/N$=5$. When we co-add three single 1200\,s-exposure-spectra for common stars and re-calculate the S/N of the co-added spectra, we find the S/N of blue band of these co-added spectra is around 10 at $G=14.5$\,mag. At $G\sim13.5$, the blue band S/N is around 20, at which we can obtain chemical elemental abundance statistically. At $G\sim12$\,mag, the blue band S/N reaches 50, which is enable to provide reliable elemental abundances for individual stars. Table~\ref{tab:Gsnr} lists the S/Ns for blue and red bands for single 1200\,s exposure and co-added $3\times1200$\,s exposure time as functions of $G$ magnitude. 

It is noted that a few spectra with smaller magnitudes have lower S/N than the majorities. This is caused that they were observed by about 20\% low-efficiency fibers. 

\begin{figure*}
\centering
\includegraphics[scale=0.5]{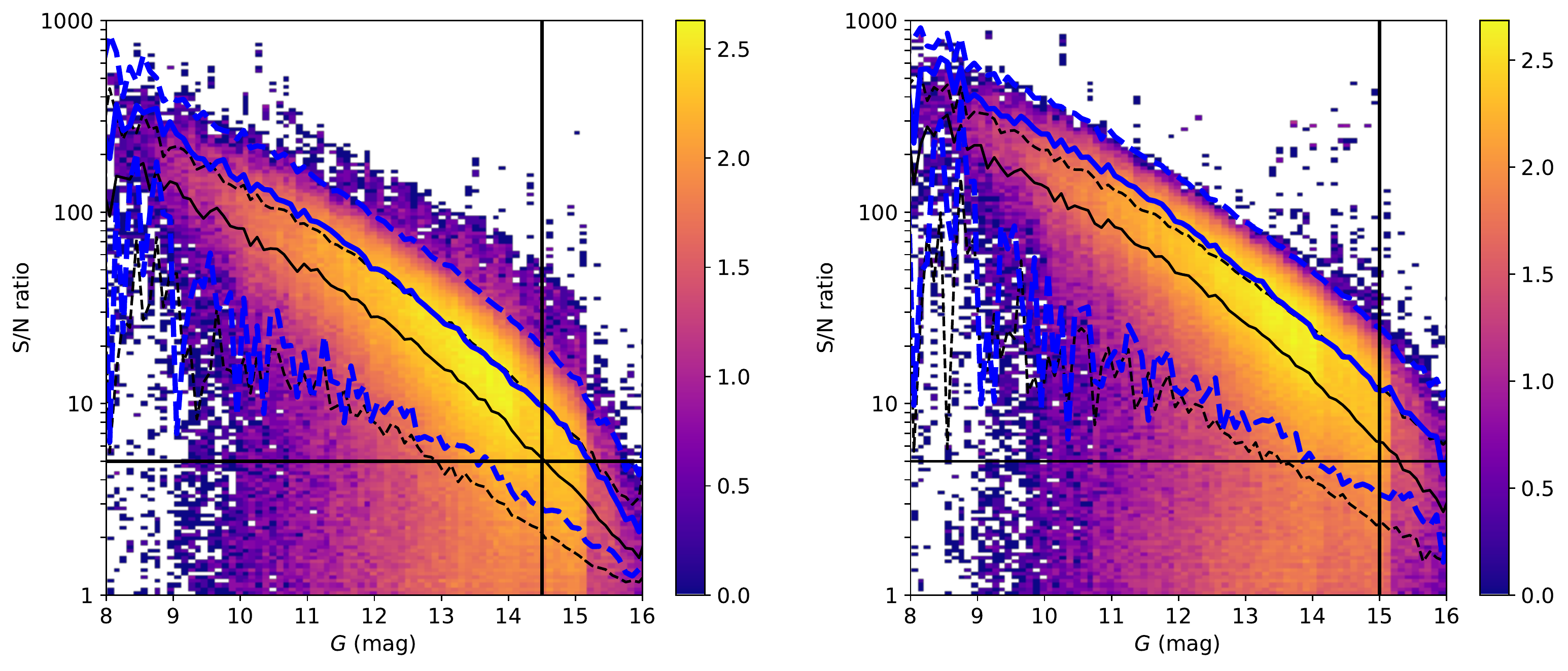}
\caption{The left panel shows the distribution of around 5 million blue-band spectra with single 1200\,s exposure in S/N (blue) vs. $G$ magnitude plane. The colors indicate the logarithmic number of stars in each bin. The black solid and dashed lines show the median S/N and 15\% (85\%) percentiles along $G$ magnitude. The thick blue solid and dashed lines show the median S/N and its 15\% (85\%) percentiles for the co-added spectra with $3\times1200$\,s exposure time. The vertical line indicates $G=14.5$\,mag, at which the single 1200\,s exposure reaches median S/N of 5. The horizontal line indicate S/N$=5$. The right panel shows the distribution of around 5 million red-band spectra with single 1200\,s exposure in S/N (red) vs. $G$ magnitude plane. The symbols are similar to the left panel. The black vertical line indicates $G=15$\,mag, which is the limiting magnitude at S/N$=5$ at red band. }\label{fig:Gsnr}	
\end{figure*}

\begin{table}
\centering
\caption{The signal-to-noise ratio vs. $G$ magnitude.}\label{tab:Gsnr}
\begin{tabular}{c|c|c|c|c}
\hline\hline
$G$ &  S/N at blue band & S/N at red band &  S/N at blue band &  S/N at red band \\
mag & 1200\,s exposure & 1200\,s exposure & $3\times1200$\,s exposure & $3\times1200$\,s exposure\\
\hline
11.0 & 51.60 & 87.47 & 96.09 & 165.53\\
11.5 & 39.12 & 65.10 & 71.92 & 120.45\\
12.0 & 28.40 & 47.83 & 50.50 & 88.69\\
12.5 & 21.73 & 36.54 & 38.72 & 66.86\\
13.0 & 15.56 & 26.25 & 27.15 & 47.64\\
13.5 & 11.13 & 19.11 & 19.48 & 34.36\\
14.0 & 7.74 & 13.61 & 13.64 & 24.05\\
14.5 & 5.18 & 9.46 & 9.75 & 17.51\\
15.0 & 3.55 & 6.30 & 6.30 & 11.78\\	
\hline\hline
\end{tabular}	
\end{table}

\cite{2019RAA....19...75L} and \cite{2019ApJS..244...27W} have investigated the performance of the radial velocity derived from cross-identification for LAMOST-MRS spectra. They illustrated that the uncertainty of the radial velocity is around 1\,\kms. 
%
%
For the binaries in time-domain spectroscopic data, in principle, we do not need the absolutely calibrated radial velocities. The relatively calibrated radial velocities should be sufficient for many of the time-domain sciences. Therefore, we are developing an independent pipeline to derive the differential radial velocity with correction of the systematic deviation occurred in each exposure. With the relatively calibrated differential radial velocities, the uncertainty of the late-type stars can reach $\sim0.2$\,\kms\ (Xiong et al. in prep), a factor of 5 better than regular radial velocity estimates. Moreover, the relative velocity calibration can also provide lots of constant stars, i.e., their velocities do not have measurable change during the observations. These constant stars are widely distributed in all time-domain fields and will be used for absolute calibration of the radial velocities used in the time-domain spectroscopic survey.  

\subsection{Survey strategy}\label{sec:sim}
According to the scientific goals, we separate the LAMOST-MRS survey into two parts---the non time-domain and the time-domain survey. The non time-domain (hereafter NT) survey will obtain a few millions of stars widely and roughly continuously distributed in the sky to incorporate with the Galactic archaeology, star forming region, nebulae area etc. Because NT survey does not use the single exposure information, we are able to co-add the single-exposure spectra for a same star to achieve higher S/N. Typically, a NT plate will continuously take 3 single 1200\,s exposure and obtain the co-added spectra with total exposure time of $3\times1200$\,s. Based on the commissioning survey, this can obtain limiting magnitude of $G=14.5$\,mag for blue band and 15\,mag for red band with S/N$\sim10$.

In the mean time, the time-domain (hereafter TD) survey focuses on some specific pencil beams and repeatedly observe the same groups of stars located in these fields over years. For TD survey, the spectra will be observed with 3--8 single 1200\,s exposures in one observation night and will be repeatedly observed many nights during the 5-year survey. For the variation of radial velocities, as we mentioned in section~\ref{sec:perf}, the single-exposure spectra will be used. Therefore, the limiting magnitude of the TD survey will reach $G\sim14$\,mag and $14.5$\,mag for blue and red bands, respectively, so that the S/N can be higher than $\sim10$. It is noted that the TD single-exposure spectra can also be co-added finally and obtained fairly high S/N for elemental abundance estimation.

Therefore, we plan to set the limiting magnitude of $G=15$\,mag for both NT and TD surveys.

As seen in section~\ref{sect:sci}, the LAMOST-MRS survey has to combine quite versatile scientific topics within one survey. We hence break down the survey into a few different sub-projects, the {\it Kepler} region survey (MRS-K), the TESS follow-up survey (MRS-T), the star forming region survey (MRS-S), the binary survey (MRS-B), the Galactic nebulae survey (MRS-N), the Galactic archaeology survey (MRS-G), and open cluster survey (MRS-O). Among them, MRS-K, MRS-T, MRS-B, and part fields of MRS-S are assigned as the TD survey, while the MRS-G, MRS-N, MRS-O, and the rest part of the MRS-S are operated as NT survey.

To determine the best ratio of TD to NT surveys, we run lots of 5-year-survey simulations with different ratios of TD to NT. In the simulations, we assume that if the field to be observed is a NT field, LAMOST will take three continuous 1200\,s exposures; if it is a TD field, LAMOST will flexibly take a few 1200\,s exposures (at least 3 times, sometimes more, depending on the specific requests of the fields) within the 4-hour observation window around meridian, which is the only sky area that LAMOST is able to observe. While the NT field is continuously distributed in the northern sky, the TD field must be pre-selected as a series of pencil beam sky area. Therefore, we randomly select 100, 200, 300, 400, and 500 TD fields, respectively, in the simulations. The simulations only assign half of a month to LAMOST-MRS survey and leave the other half for the low-resolution survey. The results of the simulations are summarized in Table~\ref{tab:simulation}. 

Within 5 years, the simulations show that LAMOST-MRS can observe around 2220 NT and TD \textit{visits} together. The term \textit{visit} is defined as following: a) if a TD field is observed in one night, no matter how many exposures it is taken, we count it as 1 visit; b) if a TD field is observed in two nights, we count it as 2 visits; c) for a NT field, a visit means one observation in a night. 

It is seen from Table~\ref{tab:simulation} that the larger fraction of the TD fields, the more observation visits for each TD field. For instance, when ${\rm TD}/({\rm TD}+{\rm NT})=20$\% and we set 100 TD fields in the simulations, we obtain that the observation visits per TD field is $701/99\simeq7.1$. However, when ${\rm TD}/({\rm TD}+{\rm NT})=$80\%, the number of visits per TD field increases to $1939/100\simeq19.39$. On the other hand, the more pre-selected TD fields, the less the number of the observed field. In other word, the more TD fields is selected, the less the completeness for each field can reach. 

Considering the performance of the orbital resolution of spectroscopic binary stars, we need to at least observe $\sim10$ visits per star. We finally adopt that the total number of TD fields is around 100 and the fraction of TD fields is 60\% among the all MRS fields by taking into account the least numbers of objects. According to the simulations, in this case, each TD field can be visited 16 times. The $\sim100$ TD fields are assigned to MRS-B, MRS-K, MRS-T, and MRS-S. Table~\ref{tab:TDfields} lists all pre-selecting TD fields with a flag of which sub-project they belong to. 

Note that the survey plan may slightly change during the 5-year survey so that the visits of observations for TD fields can be balanced to each sub-project. It is expected that, through such moderate adjustment, each sub-project can obtain sufficient epochs with proper observation intervals to satisfy its scientific goals.

\begin{table*}
\centering
\caption{Results of the survey simulations to determine the ratio of TD to NT observations.}\label{tab:simulation}
\begin{tabular}{l|l|c|c|c|c}
\hline\hline
\multicolumn{2}{l|}{Visits of TD }&\multicolumn{4}{c}{TD/(TD+NT) in visits}\\
\hline
& & 20\% & 40\% & 60\% & 80\%\\
\hline
100& No. of observed fields & 99 & 100 & 100 & 100\\
&Observation visits & 701 & 1184 & 1609 & 1939\\
\hline
200 & No. of observed fields & 192 & 195 & 200 & 199\\
&Observation visits & 707 & 1202 & 1601 & 1936 \\
\hline
300 & No. of observed fields & 264 & 291 & 295 & 298\\
& Observation visits & 707 & 1199 & 1617 & 1954\\
\hline
400 & No. of observed fields & 316 & 368 & 388 & 389\\
& Observation visits & 707 & 1199 & 1615 & 1943\\
\hline
500 & No. of observed fields & 373 & 440 & 466 & 477 \\
& Observation visits & 710 & 1196 & 1609 & 1951 \\
\hline\hline
\end{tabular}	
\end{table*}

\section{Detailed survey plan}
\label{sect:survey}
\begin{figure*}
	\centering
	\includegraphics[scale=0.8]{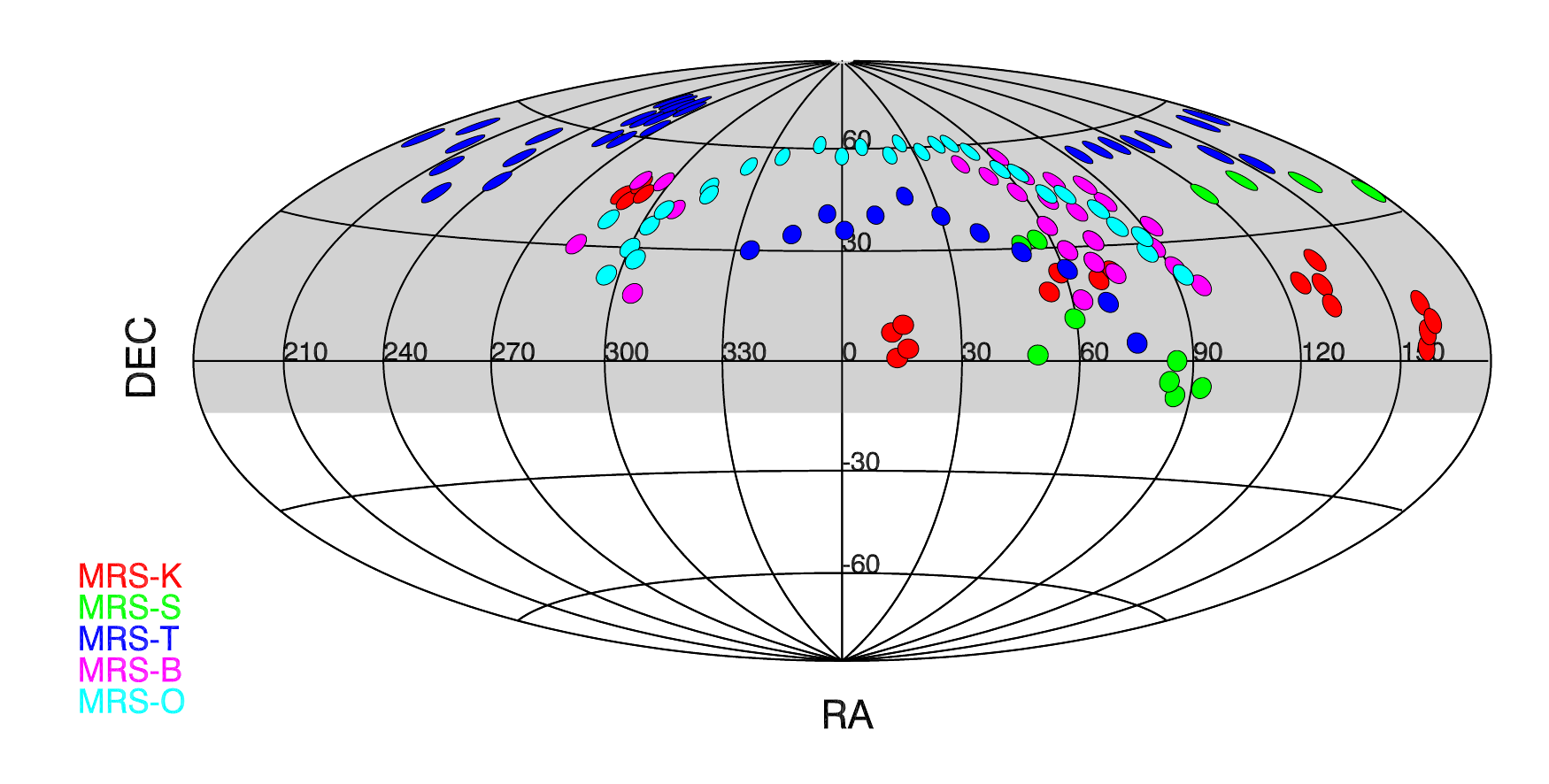}
\caption{The aitoff map shows the footprints of the LAMOST-MRS survey. The gray area indicates the NT footprint. }\label{fig:footprintTD}
\end{figure*}

\subsection{Input catalogs}\label{sec:inputcat}
In order to provide a homogeneous input catalog, the LAMOST-MRS survey uses \emph{Gaia} DR2 (\cite{2018A&A...616A...1G}) as the basic input catalog. Most of these targets are found in \emph{Gaia}. For some special targets which may not be provided based on \emph{Gaia} data, we cross-identify their \emph{Gaia} counterpart. An exception is the targeting for MRS-N, which plans to observe non-stellar objects. The sub-project sets up its own method for targeting (see Wu et al. in prep.).

\subsection{Targeting strategies}\label{sec:footprints}
Figure~\ref{fig:footprintTD} displays the footprints for the LAMOST-MRS survey. Detailed discussions about the targeting strategies for different sub-projects are in the next sections.

\begin{table*}
	\caption{MRS TD fields}\label{tab:TDfields}
\begin{center}
\scalebox{0.7}{
\begin{tabular}{c|c|c|c|c|c|c|c}
\hline\hline
Plate Name       & RA       & dec     & Type & Plate Name       & RA       & dec     & Type \\
\hline
TD005004N074006K &  12.5179 &  7.6685 & K & TD005501N004722K &  13.7558 &  0.7896 & K   \\
TD010142N094445K &  15.4287 &  9.7460 & K  & TD010605N031628K &  16.5241 &  3.2747 & K \\
TD033722N181216K &  54.3451 & 18.2047 & K & TD035321N230725K &  58.3382 & 23.1236 & K  \\
TD043446N210613K &  68.6954 & 21.1037 & K & TD045334N231856K &  73.3921 & 23.3157 & K   \\
TD082325N180811K & 125.8576 & 18.1364 & K &TD084806N172341K & 132.0265 & 17.3948 & K   \\
TD084844N123545K & 132.1871 & 12.5960 & K  & TD085754N225914K & 134.4779 & 22.9875 & K  \\
TD103356N023723K & 158.4834 &  2.6231 & K  & TD103827N055449K & 159.6151 &  5.9136 & K \\
TD104037N120443K & 160.1555 & 12.0787 & K & TD104844N081314K & 162.1836 &  8.2207 & K  \\
TD190808N440210K & 287.0347 & 44.0364 & K  & TD192102N424113K & 290.2617 & 42.6870 & K   \\
TD192314N471144K & 290.8118 & 47.1958 & K & TD193637N444141K & 294.1582 & 44.6949 & K   \\
\hline
TD031711N013329S &  49.2994 &  1.5581 & S & TD032323N305902S &  50.8497 & 30.9840 & S   \\
TD034419N321717S &  56.0797 & 32.2882 & S & TD035837N110139S &  59.6556 & 11.0275 & S   \\
TD040433N355514S &  61.1407 & 35.9206 & S & TD041324N290722S &  63.3519 & 29.1229 & S  \\
TD043724N254338S &  69.3525 & 25.7274 & S  & TD044912N312614S &  72.3035 & 31.4374 & S  \\
TD053531S051602S &  83.8807 & -5.2674 & S & TD054232S000057S &  85.6372 & -0.0161 & S  \\
TD054339S085615S &  85.9138 & -8.9378 & S & TD061101S064515S &  92.7551 & -6.7542 & S   \\
TD080016N404735S & 120.0691 & 40.7931 & S  & TD091555N422558S & 138.9804 & 42.4328 & S  \\
TD103949N392416S & 159.9556 & 39.4047 & S & TD114941N345554S & 177.4238 & 34.9318 & S   \\
\hline
TD000246N354855T &   0.6917 & 35.8154 & T &
TD004049N401113T &  10.2054 & 40.1870 & T   \\
TD012220N453143T &  20.5852 & 45.5288 & T &
TD020047N393731T &  30.1984 & 39.6254 & T   \\
TD024017N343505T &  40.0735 & 34.5849 & T &
TD032020N290254T &  50.0849 & 29.0485 & T   \\
TD040421N240621T &  61.0904 & 24.1059 & T &
TD043731N150847T &  69.3832 & 15.1465 & T   \\
TD050047N043403T &  75.1969 &  4.5676 & T &
TD062018N541459T &  95.0789 & 54.2498 & T   \\
TD070113N545124T & 105.3067 & 54.8567 & T &
TD073820N551219T & 114.5837 & 55.2055 & T   \\
TD081828N543349T & 124.6203 & 54.5638 & T &
TD085722N545046T & 134.3433 & 54.8462 & T   \\
TD094225N484633T & 145.6042 & 48.7761 & T &
TD101631N450236T & 154.1301 & 45.0435 & T   \\
TD110230N550142T & 165.6279 & 55.0286 & T &
TD114018N551809T & 175.0786 & 55.3027 & T   \\
TD121948N485902T & 184.9529 & 48.9841 & T &
TD125617N540558T & 194.0717 & 54.0997 & T   \\
TD134023N503110T & 205.0968 & 50.5194 & T &
TD141602N453358T & 214.0100 & 45.5663 & T   \\
TD145924N393841T & 224.8540 & 39.6449 & T &
TD151257N560247T & 228.2412 & 56.0465 & T   \\
TD153834N502523T & 234.6425 & 50.4233 & T &
TD160011N441646T & 240.0493 & 44.2796 & T   \\
TD161932N702206T & 244.8867 & 70.3685 & T &
TD164021N701415T & 250.0889 & 70.2376 & T   \\
TD164055N643520T & 250.2297 & 64.5890 & T &
TD165902N691110T & 254.7607 & 69.1863 & T   \\
TD165950N582745T & 254.9587 & 58.4626 & T &
TD170215N643602T & 255.5655 & 64.6007 & T   \\
TD171857N704716T & 259.7396 & 70.7878 & T &
TD172042N653823T & 260.1753 & 65.6400 & T   \\
TD172604N583906T & 261.5202 & 58.6519 & T &
TD174016N690900T & 265.0704 & 69.1503 & T   \\
TD174903N624755T & 267.2661 & 62.7988 & T &
TD221648N300528T & 334.2007 & 30.0913 & T   \\
TD230151N343645T & 345.4644 & 34.6125 & T &
TD234129N403316T & 355.3734 & 40.5545 & T   \\
\hline
TD030224N542147B &  45.6017 & 54.3633 & B  &
TD033106N502853B &  52.7775 & 50.4816 & B   \\
TD035400N453042B &  58.5035 & 45.5118 & B  &
TD040433N355514B &  61.1407 & 35.9206 & B   \\
TD041055N155648B &  62.7299 & 15.9467 & B  &
TD041114N555437B &  62.8103 & 55.9105 & B   \\
TD041324N290722B &  63.3518 & 29.1229 & B  &
TD042145N500206B &  65.4395 & 50.0352 & B   \\
TD042649N425435B &  66.7056 & 42.9099 & B  &
TD043724N254338B &  69.3524 & 25.7274 & B   \\
TD044912N312615B &  72.3035 & 31.4376 & B  &
TD045227N391703B &  73.1158 & 39.2842 & B   \\
TD045606N223436B &  74.0275 & 22.5767 & B  &
TD045704N475243B &  74.2692 & 47.8788 & B   \\
TD053403N461112B &  83.5141 & 46.1869 & B  &
TD053908N412131B &  84.7863 & 41.3589 & B   \\
TD055633N285632B &  89.1407 & 28.9423 & B  &
TD060648N233818B &  91.7027 & 23.6386 & B   \\
TD061044N341537B &  92.6860 & 34.2604 & B  &
TD062610N184526B &  96.5430 & 18.7573 & B   \\
TD185549N301843B & 283.9552 & 30.3121 & B  &
TD191631N482124B & 289.1298 & 48.3568 & B   \\
TD195013N482329B & 297.5549 & 48.3916 & B  &
TD202021N174734B & 305.0892 & 17.7930 & B   \\
TD203106N405128B & 307.7787 & 40.8579 & B   \\
\hline\hline
\end{tabular}}
\end{center}
\end{table*}

\subsubsection{MRS-B fields}
In order to evaluate binary fraction and binary parameters (e.g. orbital period, eccentric, and mass function and so on) accurately, we hope to observe each fields with 16 visits. In the MRS-B sub project, the binaries with periods from a few ten days to several hundred days are paid more attentions, so we try to make sure the observe cadence would larger than a few days and only exposure 3 times, which likes the NT observation projects, within a visit. A fixed cadence could lead to a selection effect in periods. So we do not fix the observation cadence. In addition, we randomly select two fields (TD060648N233818B and TD203106N405128B) that will  observe at least one visit in each available month of the future five years observation. In this way, we can obtain more than 16 radial velocities for each source. It would help us to get hundreds of binary parameters with periods around a year. 

\begin{figure}
    \centering
    \includegraphics[scale=0.6]{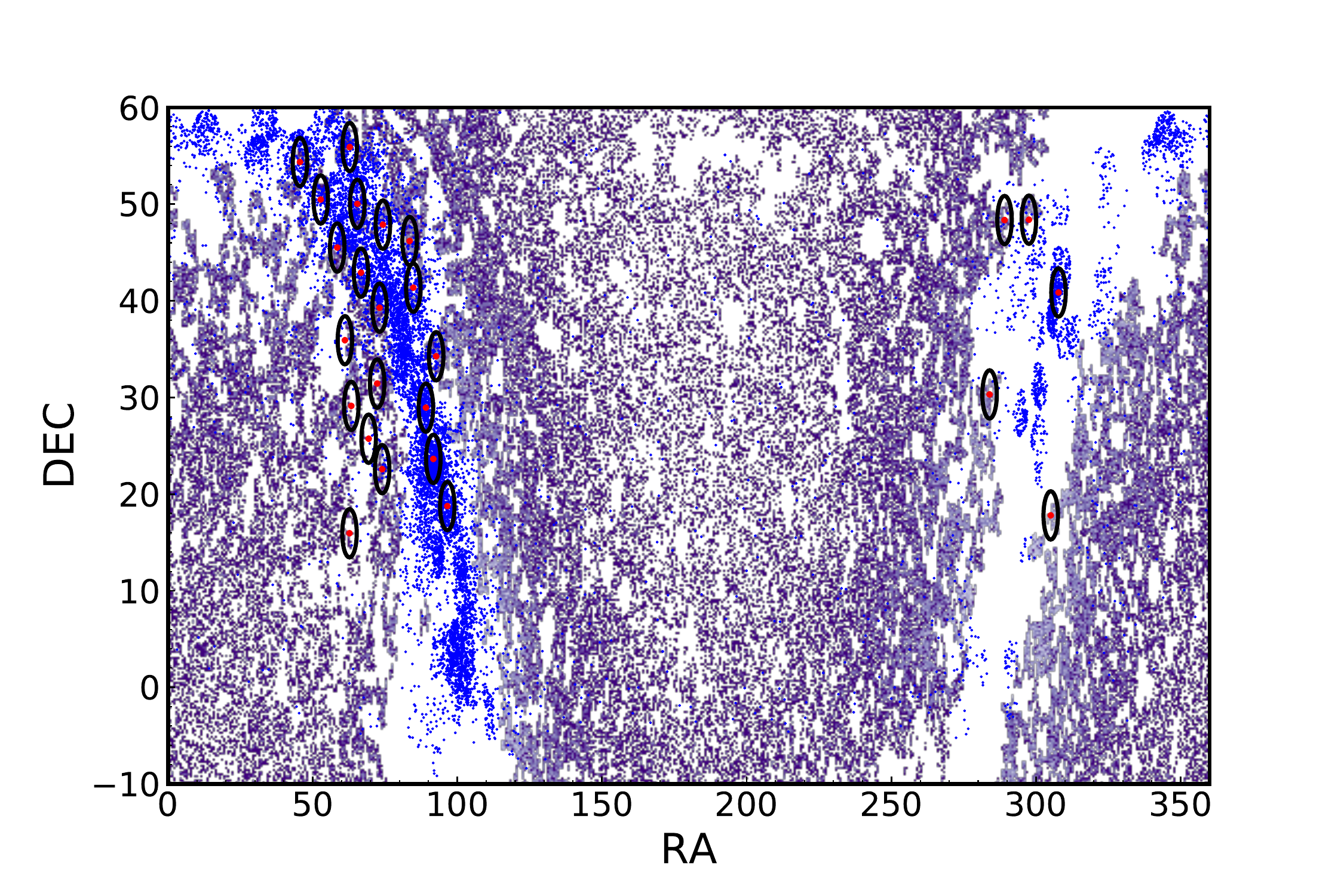}
    \caption{25 plates (black oval circles) with center stars (red solid dots) and four specific plates (red circles) for MRS-B sub-project. The blue dots represent the early stars with O or B spectral types and the purple dots denote the stars with GALEX observations.}
    \label{fig:binarity1}
\end{figure}


\subsubsection{MRS-K fields}

\begin{figure}[!htp]
    \centering
    \includegraphics[width=\textwidth]{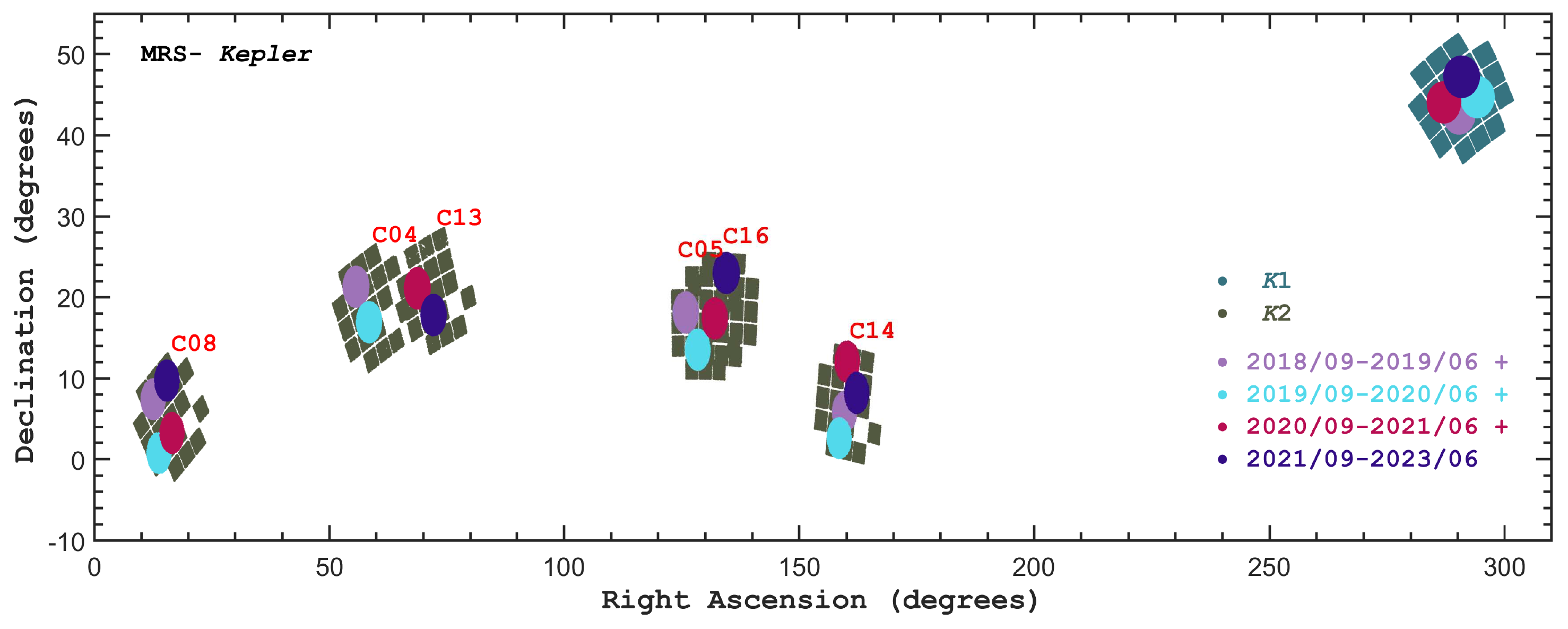}
    \caption{The spatial distribution of the 20 plates selected for MRS-K sub-project in the {\it Kepler} fields. They are divided into 4 groups, with each contains 5 plates, each of which will be observed in different years (marked in different colors).}
    \label{fig:lk}
\end{figure}

As can be seen from Figure\,\ref{fig:lk}, we plan to have LAMOST to make time-domain spectroscopic observations for 20 plates distributed in the {\sl K}1 field and four fields of the {\sl K}2 mission, i.e. the C08, C04/C13, C05/C16, and C14, with four plates in each field, respectively. On average, each plate will get around 60 exposures during the second five years of LAMOST surveys.

With the observation strategy of the MRS-K sub-project, LAMOST is expected to observe two plates in two neighboring fields in right ascension sequentially with one plate observed continuously for four hours when the plate is passing around the meridian during each night. Typically, two successful visits of LAMOST to each of the two plates in each month are needed while each plate is on the target list for 3 months with around 6 visits for one survey year. In the following survey year, the same plates are re-visited one more time and then move to the next plates in the same \emph{Kepler} fields and so on. In this way, each plate is observed for around eight times, each of which can continuously monitor the field for maximally 4 hours. In total, around 40,000 stars in the 20 plates distributed among the 5 fields of the {\it Kepler} mission will be observed. Each star is expected to have around 60 measurements with two visits per month during the 3 months observations within a year with additional one more visit in the following year. This enables one to search for both the short- and long-term radial velocity variations of stars. Once the plate had been visited at about 60 epochs, that plate will be removed from the entire observational plan. During the practice of observation, this rigorous observational plan might not be followed exactly due to some competition with other parallel programs. We refer one to see the details of the observational results from the first year in (\cite{Zong2020}).

As far as the priorities of the targets, the known pulsating stars, binary stars, stars with stellar activities and exoplanet-host stars will be on the list of top priorities. The fibers will be assigned to stars with available {\it Kepler} light curves in highest priority. The rest fibers are assigned to stars according to their brightness, while the OB type stars should have higher priorities.

\subsubsection{MRS-T fields}
We plan to monitor $\sim 10^5$ TESS stars over the course of $\sim 5$\, years with LAMOST-MRS to constrain the presence of binary companions. Furthermore, for a significant fraction of our sample, we will also benefit from existing {\it Gaia} RV measurements to extend the accessible binary period range. LAMOST-MRS and TESS synergy will potentially reveal many planet-bearing binary stars and yield the frequency of exoplanets with binary companions, shedding new light into planet formation and evolution in binaries.

\subsubsection{MRS-S TD fields}

\begin{figure}
    \centering
    \includegraphics[scale=0.65]{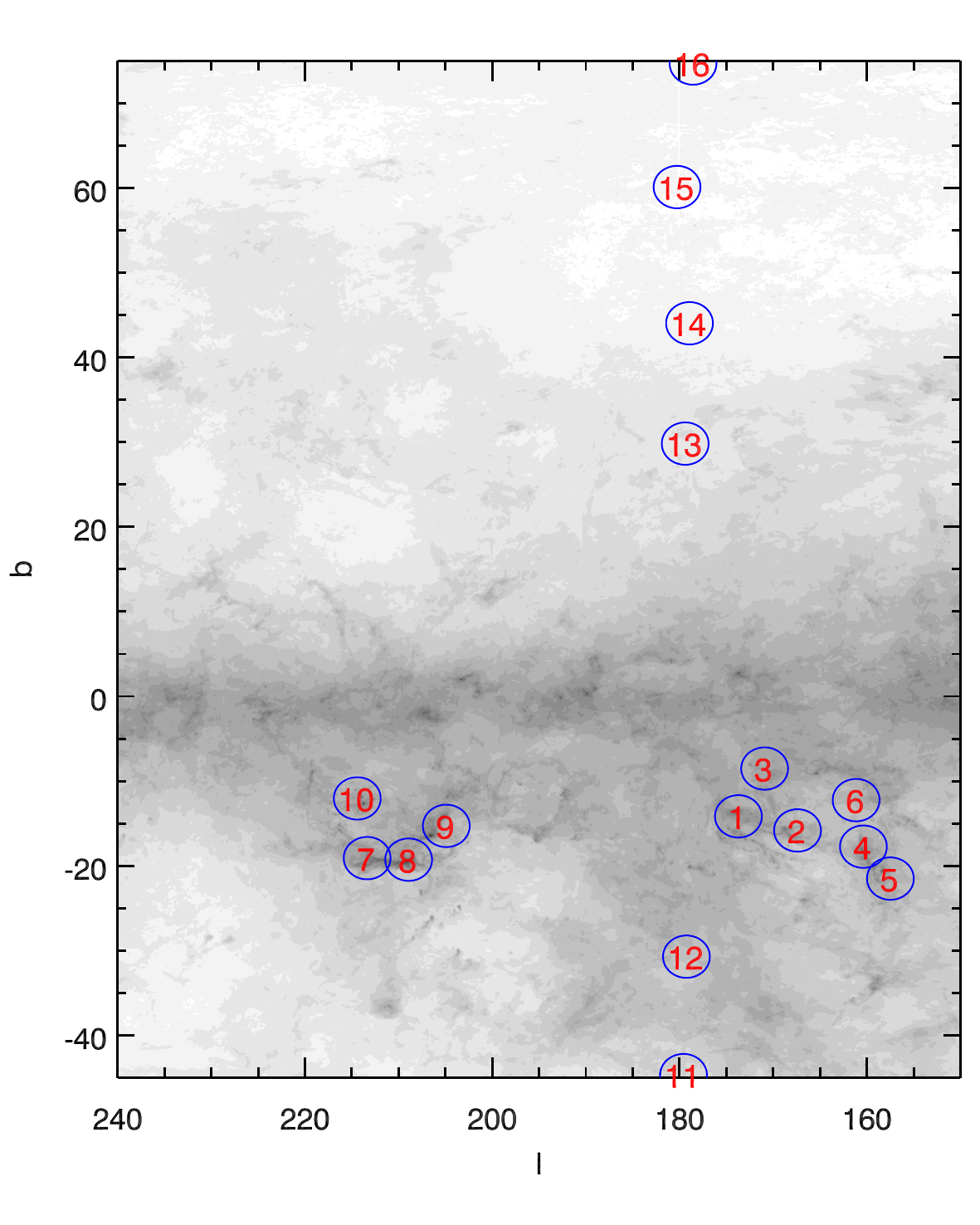}
    \caption{The circles indicate the planning fields for the MRS-S TD survey. The gray background displays the dust map of the region. The figure is drawn in Galactic coordinates.}
    \label{fig:MRS_ST}
\end{figure}

To identify binary stars in very young stellar populations, we plan to operate time-domain spectroscopic survey in the nearby star forming regions. However, because the large area of these regions, we can only select a few fields so that we can achieve the survey within 5 years. Figure \ref{fig:MRS_ST} shows the location of the selected fields with the interstellar dust map as background. The field numbers from 1 to 6 covers the Taurus and Perseus star forming regions. The numbers from 7 to 10 covers part of the Orion region. And the field from 11 to 16 are located at different Galactic latitudes so that they can provide control samples with different stellar populations. Identified young pre-main sequence stars located within these fields are marked as highest priority so that MRS-S survey can have more targets for the young stars. 

The strategy of the time-domain observations for the MRS-S TD fields is the same as that of the MRS-B fields.

\subsubsection{MRS-N fields}
From commissioning observations, we have concluded that MRS-N is seriously affected by the moonlight. So all the observation time of MRS-N should be assigned into a time range when the altitude of moon is lower than -1$^\circ$. 

The survey areas of MRS-N include two parts, the Galactic plane (GP) areas and four specific areas (see Fig.~\ref{fig:MRS-N}). The GP areas cover the northern plane within the Galactic longitude range $40^\circ < l < 215^\circ$ and the latitude range $-5^\circ < b < 5^\circ$, about 1700 square degrees. Four specific areas are Westerhout 5 (HII region), Rosette Nebula+NGC2264 (HII region), Cygnus Loop (SNRs) and Simeis 147 (SNRs), respectively. 

\begin{figure}
    \centering
    \includegraphics[width=\textwidth]{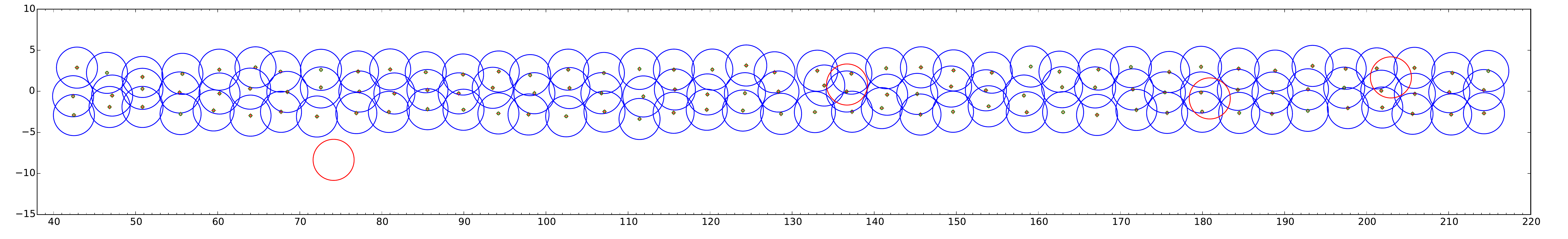}
    \caption{123 GP plates (blue circles) with center stars (red plus) and four specific plates (red circles) for MRS-N sub-project. The $x$-axis is Galactic longitude and $y$-axis is Galactic latitude.}
    \label{fig:MRS-N}
\end{figure}

We plan to use 123 plates to cover the GP areas.  The exposure time of each plate is 3$\times$900 seconds. For each specific area, the exposure time is designed as 16 $\times$ 900 seconds. So it will take about 130 hours to finish the whole MRS-N survey. MRS-N will be assigned 30 hours per year. So we can finish the MRS-N survey within 5 years without considering other objective factor such as weather. 

Because MRS-N mainly focus on the spectra of non-point source (such as the structure of nebulae), the input catalogs should be made independently unlike other LAMOST-MRS sub-projects. The input catalogs of four specific areas are man-made by comparing with optical images. From optical images, some bright stars, some sky areas and the positions of interesting structure of nebulae are selected to compose input catalogs. While the input catalog of GP area is very complex. Here we divide GP area into about 1.5 million 2 $\times$ 2 arcmin grids. Each grid can be considered as the moving area of each fiber. Then we cross-match the 1.5 million grids with Hipparcos main catalog and Pan-STARRS1 (PS1). If a star within a grid is brighter than 13 magnitude, then the star is selected as location of fiber with objtype STAR. If a star within a grid is bright than 18 and fainter than 13, then the location of fiber within the same grid should avoid the star. Else if there is not a star in a grid or the star is too faint to observe, then the location of fiber should be pointed to the brightest position within the grid.

\subsubsection{MRS-O fields}

\begin{figure}
    \centering
    \includegraphics[width=\textwidth]{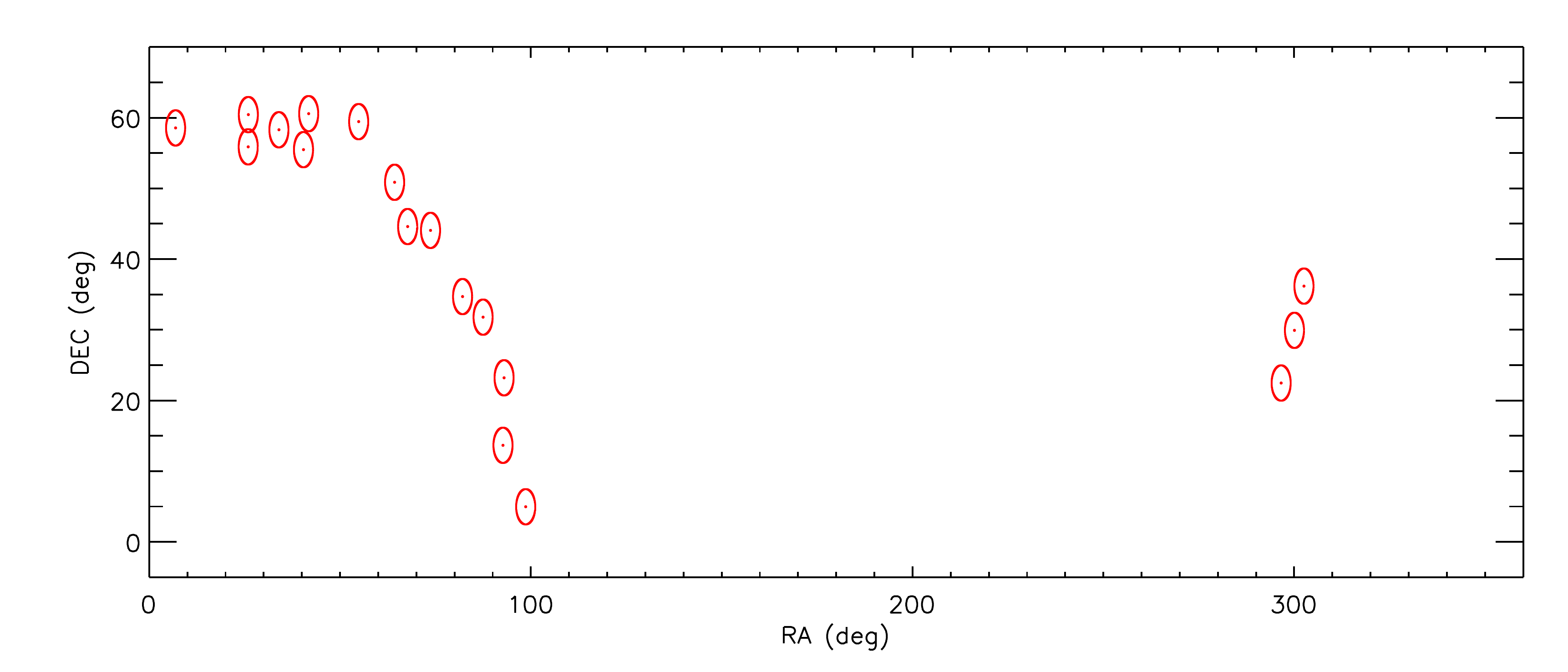}
    \caption{18 MRS-O fields with center stars (red dots) in celestial coordinates (Right Ascension and Declination). Fields are designed to cover OCs and members as many as possible.}
    \label{fig:MRS-O}
\end{figure}

In order to obtain a large number of spectra of cluster members in the LAMOST-MRS, we selected 18 open cluster related fields (hereafter OCs fields, as seen in Fig.~\ref{fig:MRS-O} which contain several tens of OCs with as many members as possible to perform the MRS-O spectroscopic observation. For each OCs field, we arranged 8 plates for observation with same center star but different fiber positions to increase the completeness of observed members. The total number of MRS-O plates is 144. Similar with other NT surveys, the exposure time of each MRS-O plate is 3  $\times$ 1200 s. It is noted that the completeness of obtained cluster members will increase to 80\% through 8 plates of spectroscopic observations, down to G=15mag.  

Since most of OCs are located on the Galactic plane, our 18 OCs fields are mainly arranged on the low Galactic latitude and along the Galactic longitude as evenly as possible. Figure~\ref{fig:MRS-O} shows the spatial distribution of the selected OCs fields. Considering  the observation mode of LAMOST, most of MRS-O plates will be observed during the period from September to February. 
    
\subsubsection{MRS-G fields}
In general, MRS-G sub-project does not have special sky area as high priority, but continuously cover the all available sky as LAMOST low-resolution survey has been doing. The limiting magnitude of the targets for MRS-G is $G=15$\,mag. The selection function of the targets is in principle the same as the previous low-resolution survey (\cite{2012RAA....12..755C,2017RAA....17...96L}).

{Same as other input catalog of the other LAMOST MRS fields, most of the targets in MRS-G filed are selected from the \emph{Gaia} DR2 catalog. We use the `square degree' as the basic block for evaluating the upper-limit of the source density in the sky. Considering a total of 4,000 fibers and a field-of-view (FOV) of 20 square degrees for LAMOST, we set an upper-limit source density of 500 stars per square degree for MRS-G field. This means for each sky area of 20 square degrees, there will be approximately 10,000 targets of maximum to be observed, which can be arranged within three plates.}

\begin{figure}
	\centering
	\includegraphics[width=\textwidth]{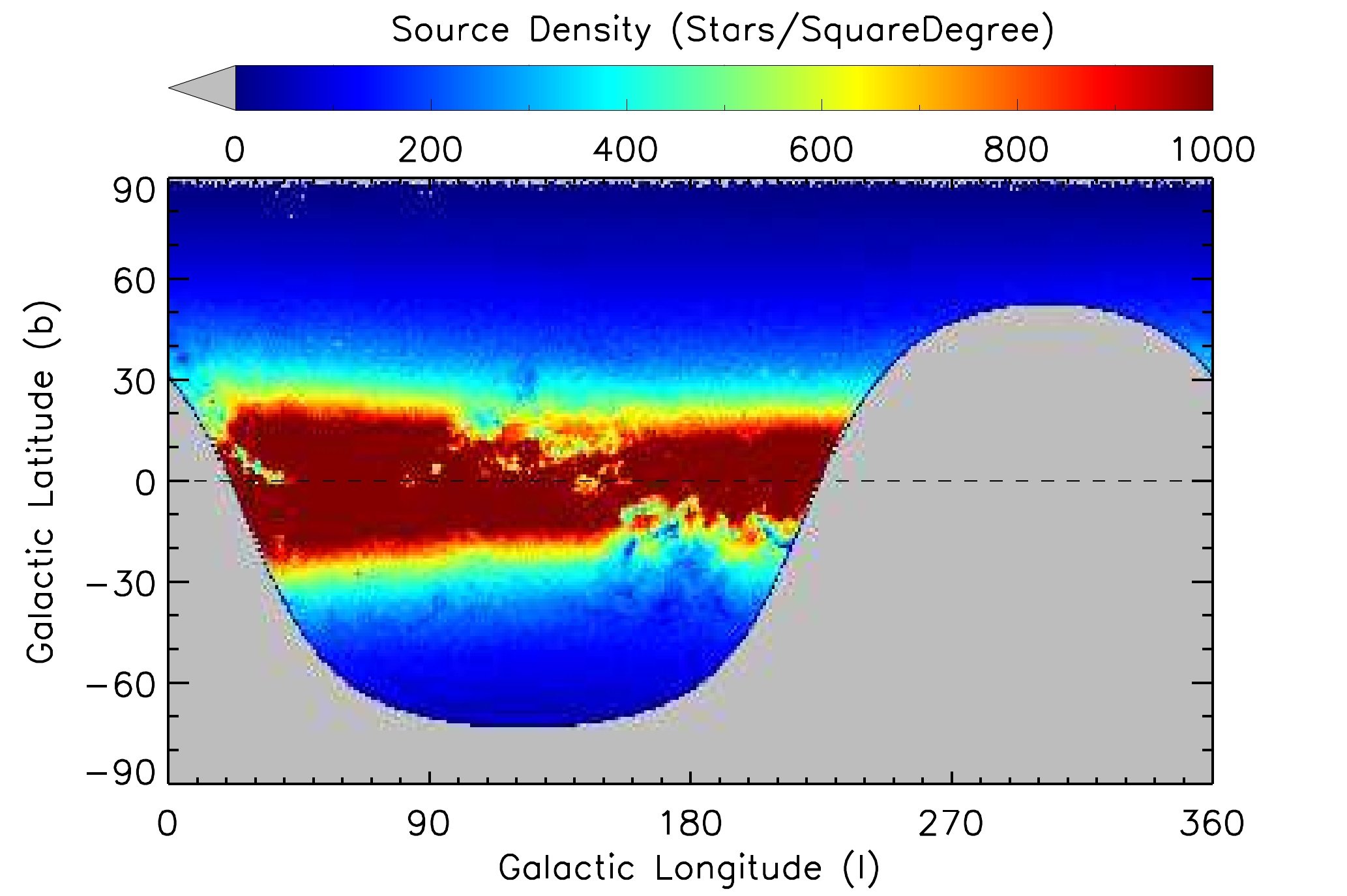}
	\caption{The number density of stars with \emph{G} magnitude between 9.0 to 15.0 mag in a Galactic coordinates map. Colors indicate different density, as illustrated on top of the figure. Although the Galactic plane is much more dense with stars, we set the upper-boundary of the density as 1,000 for a better visual effects.}
	\label{MRS-G-fig1.eps}
\end{figure}

{In Figure~\ref{MRS-G-fig1.eps}, we show the number density of the stars with \emph{G} magnitude between 9.0 and 15.0 mag in the \emph{Gaia} catalog. It is clear that the density of stars varies with the Galactic latitude. For a large fraction area of the sky, the density is below 500 stars per square degree, in such sky areas, all the targets were selected into our input catalog. While for the  higher density regions, a series criterions were applied to generate 500 stars as the sources for the input catalog. In general, we use the color of \emph{bp}$-$\emph{rp} as the filter. We first generate a grid in the dimensions of \emph{G} magnitude v.s. \emph{bp}$-$\emph{rp}. The step in \emph{G} magnitude dimension is always 0.5 mag, while for the \emph{bp}$-$\emph{rp} dimension, the steps are set as follows:}
\begin{itemize}
	\item[1)] $\emph{bp}-\emph{rp} < 0.0$ mag: treated as a single step
	\item[2)] $ 0.0 \le \emph{bp}-\emph{rp} < 6.0$ mag: 0.5 mag for each step
	\item[3)] $ \emph{bp}-\emph{rp} \ge 6.0$ mag: treated as a single step
\end{itemize}
{Thus, we got a gird with $12 \times 14 = 168$ bins in the \emph{G} v.s. \emph{bp}$-$\emph{rp} space. For the stars in the bins of $\emph{bp}-\emph{rp} < 0.0$ and $\emph{bp}-\emph{rp} \ge 6.0$, we consider them as stars of `color peculiarity', and they were always selected into our `pool' of 500 stars. While for the other bins, one target is randomly selected from each bin for each time to be filled into the `pool', and this procedure continually repeats until there is not enough space for filling with one star from each bin. Then we fill the `pool' to 500 stars with the targets selected from random, non-repeating bins throughout the grid. The selection function thus can be corrected given the known number of stars for each color bin. Figure~\ref{MRS-G-fig2.eps} shows an example of target selecting for an over-dense area of one square degree sky. The number density of our final input catalog of MRS-G is shown in Figure~\ref{MRS-G-fig3.eps}.}

\begin{figure}
	\centering
	\includegraphics[scale=0.6]{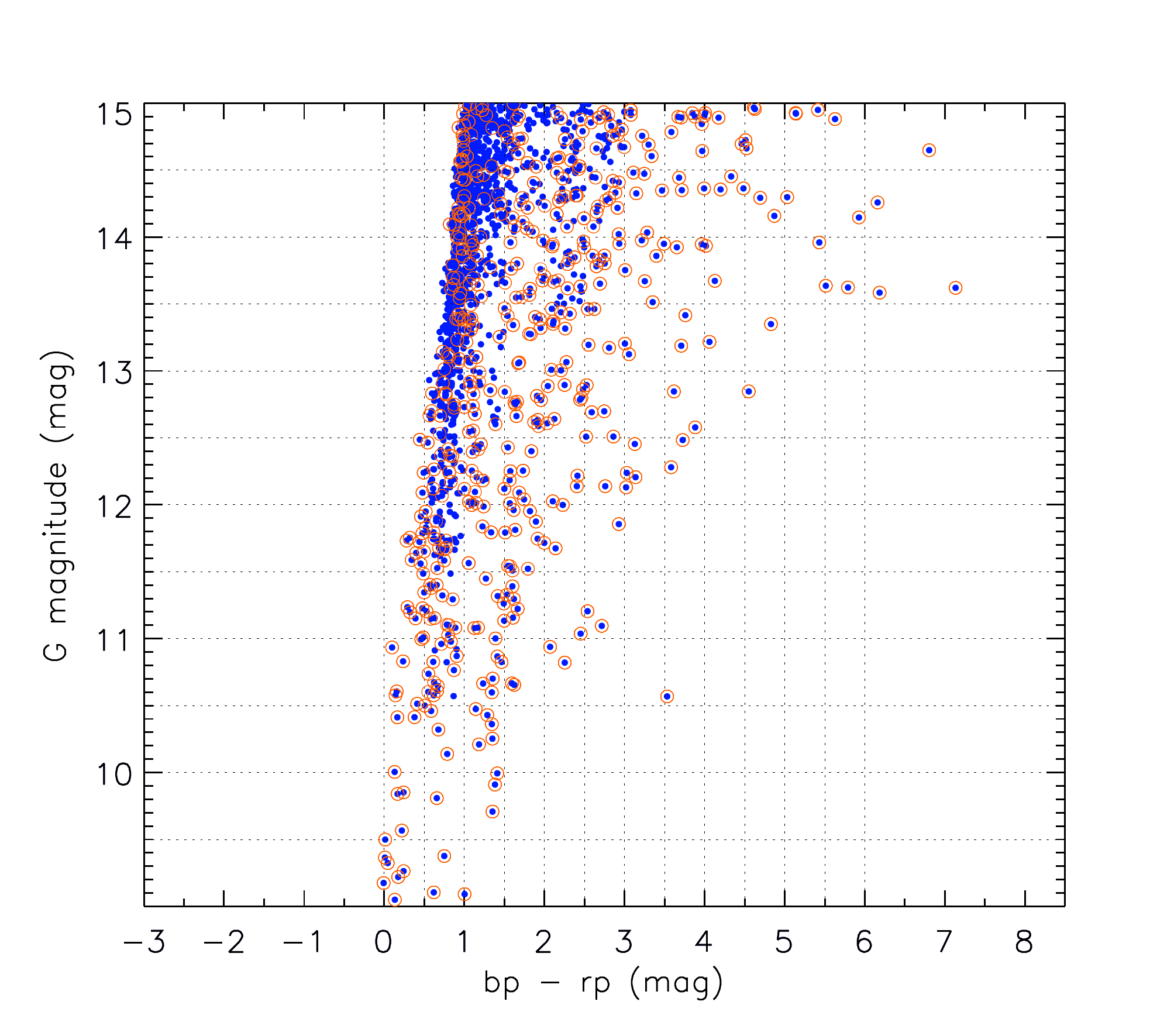}
	\caption{An example of target selecting for an over-dense area of one square degree sky. This is an area near the Galactic plane with a total of $\sim 1,500$ stars. The blue dots indicate all the stars in this area, and the red circle indicate which stars are selected. The gird is shown with doted lines. The criterion of selection ensures that the stars that are bluer or redder than the majority will be always selected into our catalog.}
	\label{MRS-G-fig2.eps}
\end{figure}

\begin{figure}
	\centering
	\includegraphics[width=\textwidth]{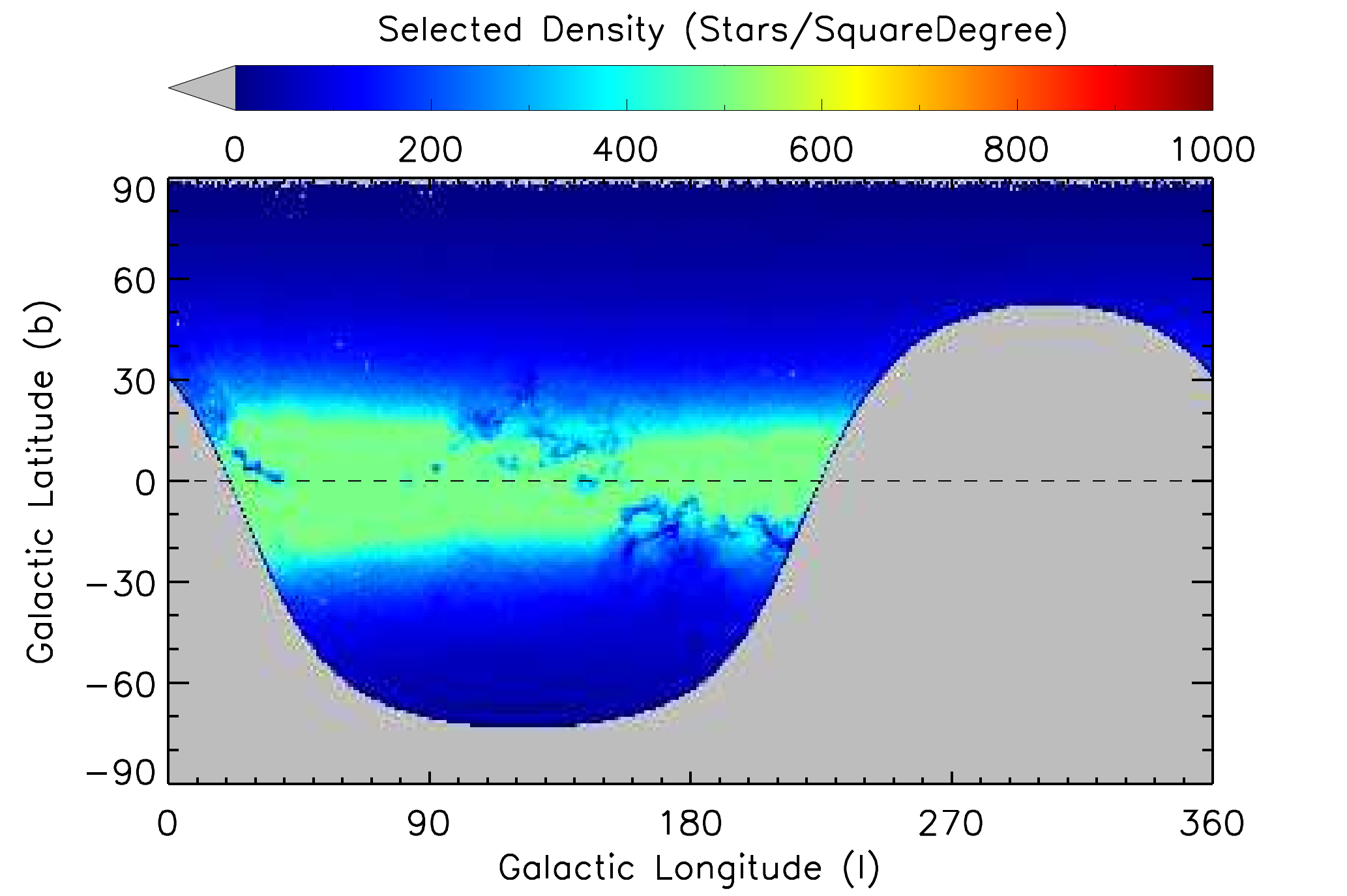}
	\caption{The number density of targets in MRS-G input catalog in a Galactic coordinates map. Colors are same as Figure~\ref{MRS-G-fig1.eps}.}
	\label{MRS-G-fig3.eps}
\end{figure}

\section{Summary}
\label{sect:summary}
As one of the largest spectroscopic survey over the world, LAMOST started to conduct medium-resolution spectroscopic survey and will significantly increase the number of stellar spectra with medium-resolution. In general, medium-resolution stellar spectra can provide more precise radial velocity as well as more information in the spectra, such as rotational velocity, elemental abundance, emission line profile etc. These improvements can support studies on binary stars, variable stars, open clusters, young stars, planet host stars, Galactic archaeology, emission nebula etc. Thus, LAMOST-MRS is able to expand the sciences from the Milky Way to more about the stellar physics and nebulae physics. Moreover, LAMOST-MRS will conduct time-domain spectroscopic survey in its 5-year survey plan. This will provide around 200 thousand stars with averagely 60 exposures. Such a large time-domain dataset will leverage the science about binary stars and variable stars in the next five years and may get some breakthrough in these topics.

\begin{acknowledgements}
This work is supported by National Key R\&D Program of China No. 2019YFA0405500. CL thanks the National Natural Science Foundation of China (NSFC) with grant No. 11835057. 
JRS and HLY acknowledge the supports from the Astronomical Big Data Joint Research Center, co-founded by the National Astronomical Observatories, Chinese Academy of Sciences and the Alibaba Cloud. 
CHH acknowledges the supports from The Science and Technology Development Fund, Macau SAR (file no. 0007/2019/A) and Faculty Research Grants of the Macau University of Science and Technology (program no. FRG-19-004-SSI). 
HNL acknowledges NSFC grant No. 11973049 and the Strategic Priority Research Program of Chinese Academy of Sciences, Grant No. XDB34020205. 
This work is also partly funded by NSFC grant No. 11833002, 11733006, 11903048, U1631131, 11973060, 11973001,11773009, 11733008, 11833006, 11973052.

Guoshoujing Telescope (the Large Sky Area Multi-Object Fiber Spectroscopic Telescope LAMOST) is a National Major Scientific Project built by the Chinese Academy of Sciences. Funding for the project has been provided by the National Development and Reform Commission. LAMOST is operated and managed by the National Astronomical Observatories, Chinese Academy of Sciences.
\end{acknowledgements}

\label{lastpage}

\end{document}